\documentclass[twocolumn,aps]{revtex4}
\usepackage{graphicx}
\usepackage{dcolumn}
\usepackage{bm,color,bbm}
\usepackage{amsmath}
\usepackage{amsfonts}
\usepackage{amssymb}

\usepackage{ulem}

\newcommand{\beq}{\begin{equation}}
\newcommand{\eeq}{\end{equation}}
\newcommand{\bqa}{\begin{eqnarray}}
\newcommand{\eqa}{\end{eqnarray}}

\newcommand{\smallfrac}[2]{\mbox{$\frac{#1}{#2}$}}

\newcommand{\half}{\smallfrac{1}{2}}

\newcommand{\blk}{\color{black}}

\definecolor{ngreen}{rgb}{0.2,0.6,0.2}

\definecolor{golden}{rgb}{0.8,0.6,0.1}

\begin{document}
\title{Experimental device-independent verification of quantum steering}
\author{Sacha Kocsis$^{1,2}$, Michael J. W. Hall$^1$, Adam J. Bennet$^1$, Dylan J. Saunders$^{1,3}$ and Geoff J. Pryde$^1$} 

\affiliation{${}^1$Centre for Quantum Dynamics, Griffith University, Brisbane, QLD 4111, Australia\\
 ${}^2$Institut f\"ur Gravitationsphysik, Leibniz Universit\"at Hannover and Max-Planck-Institut f\"ur Gravitationsphysik (Albert-Einstein-Institut), Callinstrasse 38, 30167 Hannover, Germany \\
 ${}^3$Clarendon Laboratory, Department of Physics, University of Oxford, Oxford OX1 3PU, United Kingdom}


\begin{abstract}
 Bell nonlocality between distant quantum systems---i.e., joint correlations which violate a Bell inequality---\blk can be verified without trusting the measurement devices used, nor those performing the measurements. This leads to unconditionally secure protocols for quantum information tasks such as cryptographic key distribution.  However, complete verification of Bell nonlocality  requires high detection efficiencies, and is not robust to the typical transmission losses that occur in long distance applications.  In contrast, quantum steering,  a weaker form of  quantum correlation,   can be verified for arbitrarily low detection efficiencies and high losses. The cost is that current steering-verification protocols require complete trust in one of the measurement devices and its operator,   allowing only one-sided secure key distribution.  
We present device-independent steering protocols that remove this need for trust, even  when \blk  Bell nonlocality is not present. We experimentally demonstrate this principle for singlet states and  states that do not violate a Bell inequality.

\end{abstract}

\maketitle

Entanglement provides a fundamental resource for a range of quantum technologies, from quantum information processing to enhanced precision measurement  \cite{mikeike, entang, ralph10,xiang11}. \blk In particular, the strong correlations inherent in shared entanglement---between two parties, for example---allows secure messaging and quantum information transfer, potentially over long distances  \cite{mikeike,gisinrev}. \blk At the same time, the strong restrictions of quantum measurement theory (on obtaining knowledge of observable properties through measurement) prevents the extraction of useful information when an adversary  has access to only one of the entangled systems  \cite{prevedel11,rozema12,weston14}.  Furthermore, \blk any adversary measuring one or more of the entangled systems reveals their presence to the communicating parties.

When  correlations  due to quantum entanglement are sufficiently strong, they allow the \textit{unconditionally secure} sharing of a cryptographic key between two distant locations, without requiring any trust in the devices used or in the observers reporting the results   \cite{devindqkd}. \blk It also allows generation of unconditionally genuine randomness, again with no trust in the devices used or their operators \cite{rand1,rand2}.  The corresponding verification protocols can be put in the form of a ``Bell nonlocal game'', played between a referee and two untrusted parties, which can be won by the latter only if they genuinely share a Bell-nonlocal quantum state (Fig.~1a) \cite{bellgame}, that is, an entangled state that violates a Bell inequality. \blk

There are, however, practical difficulties in entanglement verification via Bell nonlocal games.  
 Even if the entanglement is strong enough (compared to noise) to otherwise violate a Bell inequality,  there may be too many null measurement results for unconditional verification --- arising, for example, from detector inefficiencies or the typical transmission losses involved in implementations over long distances.  A sufficiently high proportion of null results will make it impossible even for `honest' devices to win a Bell-nonlocal game.  This is the well known ``detection loophole'' \cite{detloop}.

\begin{figure*}[!t]
\centering
\includegraphics[width=\textwidth]{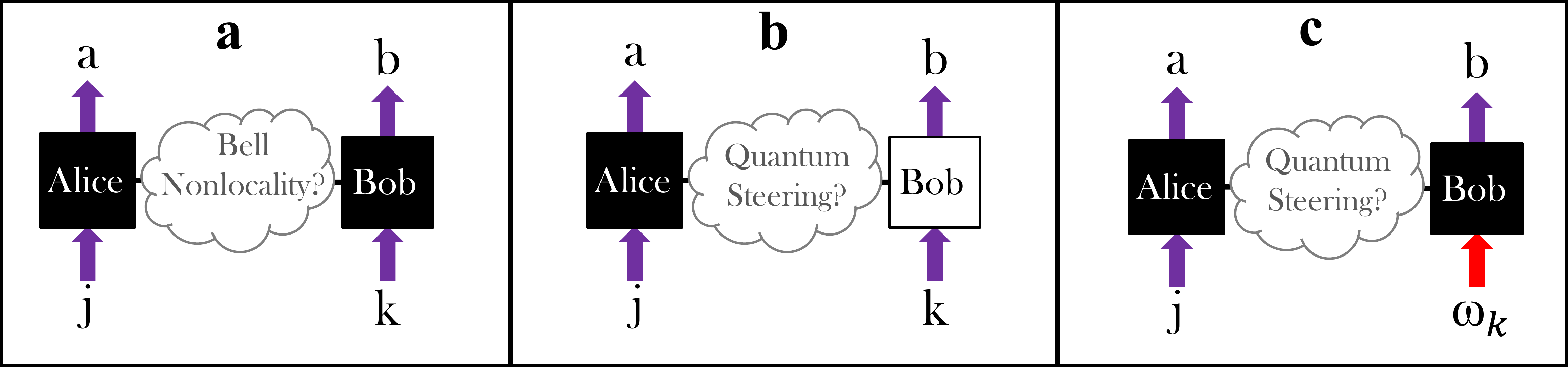}
\caption{{\bf Entanglement verification games.} {\bf (a)} In `nonlocal games'  a referee can verify that Alice and Bob share a Bell-nonlocal resource, by sending input signals $j$ and $k$, receiving output signals $a$ and $b$, and checking whether the corresponding correlations violate a Bell inequality. No trust in Alice and Bob or their devices is necessary, as indicated by the black boxes.  {\bf (b)} The referee may similarly use a `steering game' to verify the presence of a quantum steering resource, by checking whether the correlations violate a suitable steering inequality.  However, all known steering games require the referee to fully trust one of the observers and their devices, as indicated by the transparent box.  {\bf (c)} Using the device-independent protocols of this paper, the referee can now unconditionally verify  steering entanglement, by using `quantum-refereed steering games' that replace the need for trust  with   {\it quantum} input signals $\omega_k$.}
\label{fig1}
\end{figure*}

A promising alternative is based on a different test of  nonlocality,  called quantum steering  (or EPR-steering).  First identified by Erwin Schr\"odinger \cite{schr},  and present in the Einstein-Podolsky-Rosen paradox  \cite{epr},  this corresponds to being able  to use entanglement to  steer the state of a distant quantum system by local measurements, and is strictly weaker than Bell nonlocality \cite{wiseprl}. Further, the detection loophole can be circumvented in the verification of  steering,  if the device and operator for one of the two  entangled systems is completely trusted by the referee \cite{steeringdet,evans,reidpra} (Fig.~1b).  This leads to the real possibility of {\it one-sided} device-independent secure key distribution that is robust to both detector inefficiency and transmission loss \cite{onesided}.  Unfortunately, however, an unconditionally-secure protocol cannot rely on trust in even one side.  

Very recently, work on entanglement verification by Buscemi \cite{buscemi} has been generalised to show that quantum steering can in fact be verified in the absence of trust in either side, via quantum-refereed steering games \cite{cav}.  In comparison with Bell nonlocal games, the referee still sends classical signals to one party, but sends {\it quantum} signals to the other party (Fig.~1c).  The quantum signals must be chosen such that they cannot be unambiguously distinguished, to prevent the possibility of cheating.  Until now,  only an existence proof for such games  was  known, with no explicit means of construction \cite{cav}.  For the case of entanglement witnesses, a recent measurement-device-independent protocol  and   demonstration has addressed a similar question  \cite{branc, chinese},  although steering, Bell inequality violations, and calibration of the referee states (see below) were not considered. 

In this paper we give the first explicit construction of a quantum-refereed steering game, for the trust-free verification of steering entanglement.   We also  demonstrate a proof-of-principle implementation, for optical polarisation qubits, in a scenario where no Bell nonlocality ---as tested by the Clauser-Horne-Shimony-Holt (CHSH) inequality \cite{CHSH}---is  present. The results open the way to unconditionally secure key distribution protocols that do not require Bell nonlocality, and which can circumvent the detection loophole.

\section*{RESULTS}

\subsection*{Quantum-refereed steering game}

Consider the following  scenario (Fig.~1c).  On each run the referee, who we shall call Charlie, chooses at random a pair of numbers labelled by $k\equiv(j,s)$, with $j\in\{1,2,3\}$ and $s=\pm1$. Charlie sends Alice the value of $j$ as a classical signal, and sends Bob a qubit in the $s$-eigenstate of the Pauli spin observable $\sigma^C_j$, i.e., the state $\omega^C_k=\half(\mathbbm{1}+s\sigma^C_j)$. 
The referee requires Alice and Bob to send back classical binary signals, $a=\pm 1$ and $b=0$ or $1$, respectively.  
The referee uses their reported results over many runs to calculate the payoff function
\begin{equation}  \label{payoff}
P(r) :=  2  \sum_{k=(j,s)} \left[ s \langle a b \rangle_{j,s} -(r/\sqrt{3})  \langle b\rangle_{j,s}\right]  ,
\end{equation}
 where $\langle \cdot\rangle_{j,s}$ denotes the average  over those runs with $k=(j,s)$. Here $r\geq1$ is a parameter that indicates how well the referee can prepare the desired qubit states $\omega^C_k$, with $r=1$ for perfect preparation (see Methods section).  Alice and Bob win the game if and only if the payoff function is positive, i.e., if and only if $P(r)>0$.

Charlie makes no assumptions as to how Alice and Bob generate the values $a$ and $b$ on each run, as they and their devices are untrusted. Alice and Bob are told the rules of the game, and are allowed to plan a joint strategy beforehand, but cannot communicate during the game (this may be enforced by having them generate their values in spacelike separated regions, so that communication would require sending signals faster than the speed of light). 
Remarkably, despite the absence of trust by Charlie, Alice and Bob cannot cheat --- they are only able to win the game if they genuinely share quantum steering entanglement (see Methods section).

For example, suppose that Alice and Bob share a two-qubit Werner state,  $\rho^{AB}_W=W|\Psi^-\rangle^{AB}\langle\Psi^-|+(1-W)\mathbbm{1}/4$,  where $0\leq W\leq1$ and  $|\Psi^-\rangle^{AB}$  denotes the singlet state \cite{Werner89}, and adopt the following strategy: on receipt of signal $j$ Alice measures $\sigma^A_j$, while Bob measures the projection operator  $|\Psi^-\rangle^{BC}\langle\Psi^-|$  onto the singlet state in the two-qubit Hilbert space spanned by his system and $\omega^C_k$. It is straightforward to calculate
that the corresponding  theoretical  value of the payoff function in Eq.~(\ref{payoff}) is
\begin{equation} \label{pwr}
P_W(r) =  3W - \sqrt{3}r .
\end{equation}
Hence Alice and Bob can, in principle, win the game whenever $W>r/\sqrt{3}$.  This condition is in fact necessary for them to be able to win the game with a shared Werner state (see Methods section), and hence the above strategy is optimal. \blk

\subsection*{Device-independent verification of quantum entanglement}

We experimentally  
 verified device-independent EPR-steering using our quantum-refereed game. Alice and Bob's shared state, and the states sent by Charlie to Bob, were encoded in photon polarization qubit states.  The payoff function $P(r)$ was calculated via single qubit measurements and a partial Bell state measurement, all using linear optics and photon counting. \blk

In the experiment (Fig. 2), Charlie's photon source generated photon pairs that were unentangled in polarisation, and degenerate at 820 nm. One photon encoded the polarisation state $\omega_k$ and was transmitted to Bob, while the other photon acted as a heralding signal. The other photon source generated polarization entangled photon pairs at 820 nm that were shared between Alice and Bob. Alice was represented by a single qubit measurement station, used to measure one half of the entangled state. Bob was represented by a partial Bell-state measurement (BSM) device \cite{Michler96}, which performed a joint measurement on Bob's half of the entangled state and  $\omega_k$. Bob determined projections onto the singlet subspace $|\Psi^-\rangle^{BC}\langle\Psi^-|$  (corresponding to b=1) and the triplet subspace $(\mathbbm{1}-|\Psi^-\rangle^{BC}\langle\Psi^-|)$ (corresponding to b=0), of the two-qubit Hilbert space spanned by his and Charlie's systems (see Methods).

In principle, the workings of Alice's and Bob's devices need not be known,  as would indeed  be the case in a field demonstration.

 A \blk key innovation of our   protocol \blk is that the payoff function $P(r)$   (Eq.~(\ref{payoff}))  cannot present `false positives' of  steering verification. That is, Alice and Bob do not have to be trusted, and can try to cheat by any means, provided that they cannot communicate during the demonstration (this latter requirement is also necessarily mandatory in any Bell test)--- only a steerable state can ever yield a positive payoff value. It is also required that, in calculating the payoff function, Charlie chooses $r \geqslant r^*$, where $r^* (\geq 1$, $r^*=1$ perfect) characterises the quality of Charlie's preparation in the states he sends to Bob (see Methods); in this work we choose $r=r^*$. Given these conditions, robust device-independent verification of steering is possible.

 In  our test of the payoff function $P(r)$ in Eq.~(\ref{payoff}), Charlie   sent  Bob a qubit $\omega_{ k}$ (derived from the polarization-unentangled photon source)  encoded in the  $\sigma_1=\hat{X}$, $\sigma_2=\hat{Y}$, or $\sigma_3=\hat{Z}$ \blk
basis, and  announced \blk to Alice a corresponding value of $j=1$, $2$, or $3$. Alice implemented a measurement on her half of the entangled state (projective, in the $\hat{X}$, $\hat{Y}$, or $\hat{Z}$ basis depending on Charlie's announcement) and Bob implemented his partial BSM. Charlie received classical outputs from Alice ($a=\pm1$) and Bob ($b=0$ or $1$) over many runs. Using this information, Charlie calculated the payoff function   $P(r)$  \blk and tested for positivity (verifying steering). 

\begin{figure}[!t]
\centering
\includegraphics[width=0.5\textwidth]{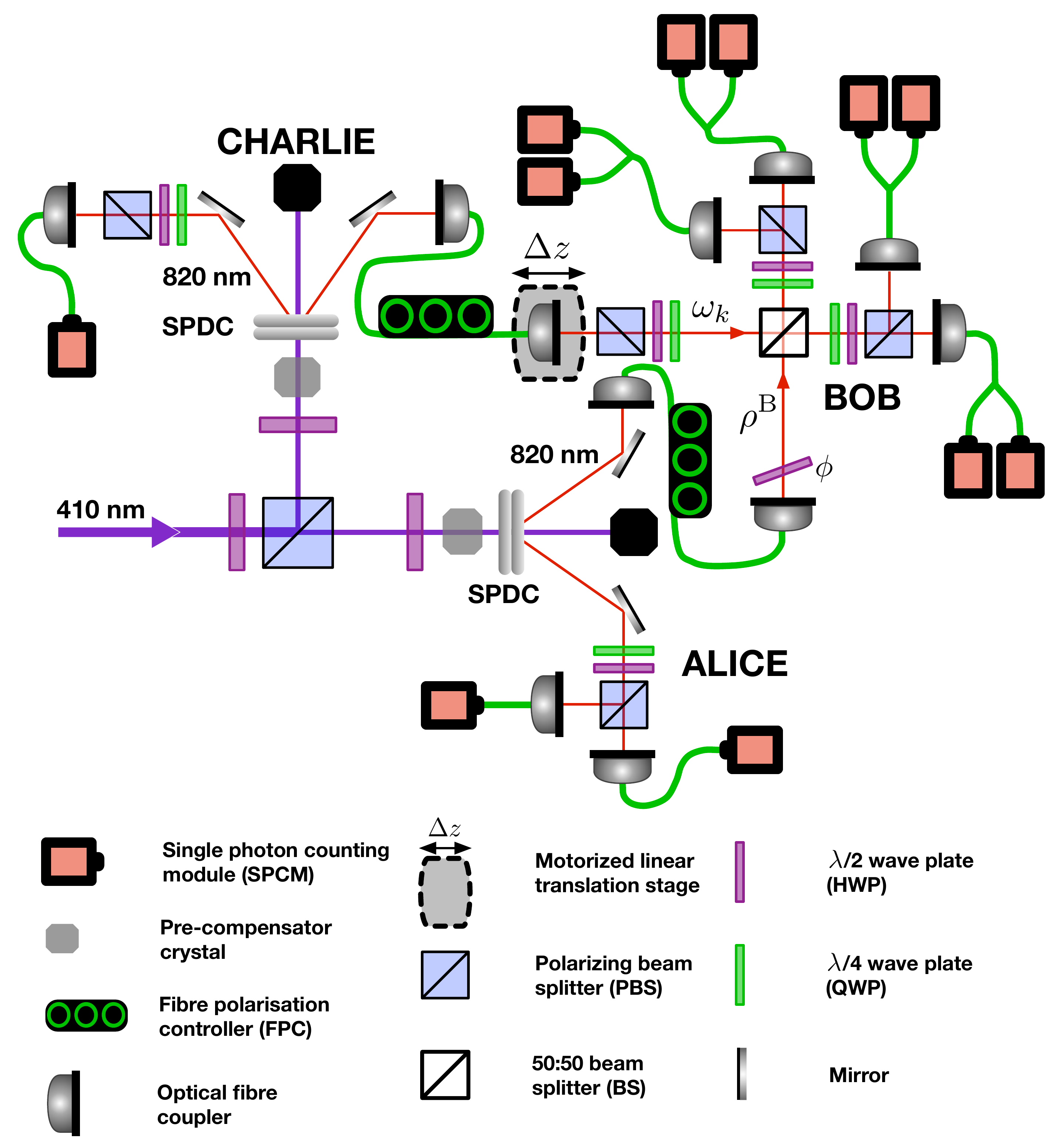}
\caption{Illustration of experimental  apparatus. \blk A pair of separate spontaneous parametric down conversion (SPDC) sources create Alice's, Bob's and Charlie's photons. One photon from Charlie's source acts as a heralding signal, with the remaining photon prepared in the quantum state $\omega_{ k}$ and sent via optical fibre to the input of Bob's partial BSM device, accompanied by a  corresponding  classical signal $j\in\{1,2,3\}$ sent to Alice. Using a 50:50 beam splitter  BS, \blk Bob combines Charlie's photon (prepared in state $\omega_{ k}$) with his own photon $\rho^{B}$ (comprising half of the entangled state $\rho^{AB}$ shared with Alice),  and projects onto the singlet subspace $|\Psi^{BC}_-\rangle\langle\Psi^{BC}_-|$. Alice receives Charlie's announcement $j$ accompanied by the other half of the  shared  entangled state   $\rho^{AB}$,  and measures $\sigma_{j}$. To execute the entanglement verification, Charlie receives Alice's and Bob's output signals $a\in\{\pm1\}$ and $b\in\{0,1\}$, and computes a payoff function $P$,  where $P>  0$  witnesses quantum  steering  in a device-independent setting.}
\label{fig2}
\end{figure}

 We tested for device-independent steering in the regime where a Bell inequality cannot be violated. In theory, the bound $P\leqslant 0$ for our steering test   requires \blk  $W>1/\sqrt{3} \approx 0.5774$  (see Methods section), \blk while the best explicit Bell-type inequality for Werner states is violated for $W \gtrsim 0.7056$---the V\'{e}rtesi bound \cite{vertesi08}---slightly below the well-known CHSH bound of $W > 1/\sqrt{2}$ \cite{CHSH}. (Note that it remains an open question whether there exists a Bell inequality that can be violated for $0.6595 \lesssim W \lesssim 0.7056$, and it is known to be impossible for $W \lesssim 0.6595$ \cite{vertesi08,acin06}).  

\begin{figure}[!t]
\centering
\includegraphics[width=0.5\textwidth]{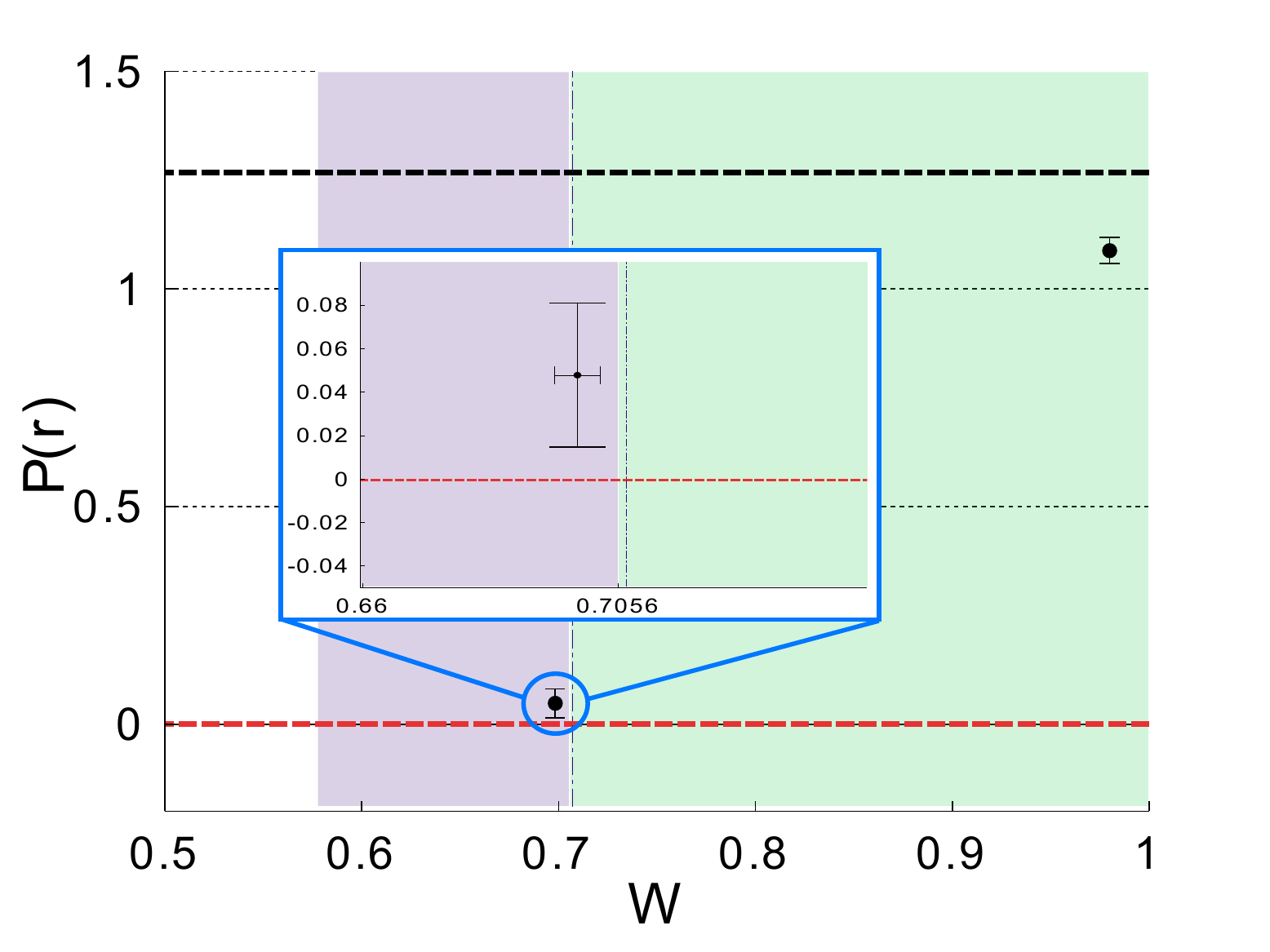}
\caption{  Observed payoff function for Werner and singlet states.  The main figure shows the measured values of the payoff function $P(r)$ for $r=r^* = 1.081 \pm 0.009$, for the cases of (i) a Werner state with $W=0.698 \pm 0.005$, and (ii) a state with fidelity $F \approx 0.98$ to the ideal singlet Bell state ($W=0.98$).  The upper dashed horizontal line indicates the maximum possible payoff, $3-\sqrt{3}$ (see text),  while the lower dashed horizontal line at $P(r)=0$ denotes the cutoff value for demonstrating steering.  The purple shaded region indicates the range of $W$ corresponding to steerable Werner states that do not violate any known Bell inequality, and the dot-dashed vertical line corresponds to the minimum value of $W$ required to violate the standard CHSH Bell inequality (see text).  As is most clearly seen in the inset figure, the data point for $W=0.698 \pm 0.005$ lies  to the left of the values required to  violate known Bell inequalities, with $P(r) > 0$. Hence steering is verified. \blk }
\label{fig3}
\end{figure}

We carefully characterised   Charlie's \blk state preparation  to determine that  $r^* = \blk 1.081 \pm 0.009$. \blk Using a Werner state with $W=0.698 \pm 0.005$ (below both the CHSH and V\'{e}rtesi bounds) we observed $P( r^* \blk)=0.05 \pm 0.04$--- a violation of our steering inequality (Fig. \ref{fig3}). This violation may be compared with the theoretical prediction $P_W( r^* \blk)=0.22$ from Eq.~(\ref{pwr}), for ideal qubit and Bell state measurements.   Thus, even without ideal measuring devices,  Charlie was able to verify that Alice could steer Bob's state, without requiring any trust in them or their devices. \blk 

With higher  values of \blk $W$ (e.g. $W \approx 1$) one would also expect a  verification of steering, \blk and indeed we observed  $P(r^*) \blk =1.09 \pm 0.03$ \blk for a state having a fidelity $F \approx 0.98$ with the ideal singlet Bell state  (Fig.~\ref{fig3}). This is close to the ideal value of $3-\sqrt{3}\approx1.13$ for a singlet state, corresponding to $W=r=1$ in Eq.~(\ref{pwr}). \blk

\section*{ DISCUSSION }

EPR-steering is a key  quantum resource  because, apart from its fundamental interest, it is known to be useful in secure quantum key distribution protocols \cite{onesided}. Compared with violation of a loophole-free Bell inequality---which provides fully device-independent QKD---EPR-steering in its usual form provides a  one-sided  device-independent protocol,  requiring trust in  one party (Bob, say) and their apparatus.  Our demonstration of  quantum-refereed steering   removes the need to trust Bob and his apparatus,  only requiring \blk the assumption that quantum mechanics is a reliable description of  reality. This lack of trust is possible essentially because Bob is unable to unambiguously distinguish between the states sent to him  by Charlie \cite{cav}. \blk

Thus, as long as quantum mechanics is correct, the protocol has the advantages of Bell inequality violation, but can tolerate higher noise (i.e.\ it works with entangled states with a higher degree of mixture).  It should also be noted that steering inequalities exist for arbitrarily high degrees of noise and loss \cite{steeringdet}, and hence corresponding quantum-refereed steering games can be constructed using our methods for  long-distance applications such as secure quantum networks \cite{network}. \blk

We note that the $r$ parameter that we  have \blk introduced is only required to characterise the degree of  confidence in the preparation of  the referee states.  It is unnecessary to characterize the state that Bob eventually receives from Charlie; \blk  indeed,  transmission through any quantum channel  will not change the protocol nor increase $r$  (see Methods section). Therefore, as long as Charlie can  characterise  his prepared states,  the protocol can proceed.   Our protocol imposes a more complex measurement procedure on Bob, a joint Bell state measurement, compared to one-qubit Pauli projections required in a Bell test. As the protocol is robust against preparation and transmission imperfections of the referee states, this added complexity of Bob's measurement is a reasonable overhead for removing all need for trust. We note that it is easier for Alice and Bob to demonstrate EPR steering to Charlie if he can prepare his states with a high degree of confidence, i.e., with $r\approx 1$. \blk

A future challenge is to demonstrate the closure of the detection loophole and spacelike-separation loophole for our protocol. When this is achieved, it will be possible to perform fully device-independent entanglement sharing between two parties---with only the assumption that quantum physics holds---with application in quantum key distribution, random number generation and beyond.   

\acknowledgments  We thank Howard Wiseman for discussions.  This work was supported by the Australian Research Council, Project No. DP 140100648.

 { \footnotesize  

\section*{ METHODS}

\subsection*{Constructing quantum-refereed steering games} 

A quantum state  $\rho^{AB}$ on some Hilbert space $H_A\otimes H_B$, shared between two parties Alice and Bob, is defined to be nonsteerable by Alice if and only if there is a local hidden state (LHS) model $\{ \rho^B_\lambda;p(\lambda)\}$ for Bob \cite{wiseprl}, i.e., if and only if the joint probability of measurement outcomes $a$ and $b$, for arbitrary measurements ${\cal A}$ and ${\cal B}$ made by Alice and Bob, can be written in the form
$p(a,b) = \sum_\lambda p(\lambda) p(a|\lambda) p(b|\lambda)$,
with $p(b|\lambda)$ restricted to have the quantum form ${\rm Tr}_B[\rho^B_\lambda {\cal B}_b]$. 
Here $\{ {\cal B}_b\}$  is the positive-operator-valued measure (POVM) corresponding to ${\cal B}$.    Local hidden state models, and hence nonsteerable states,  satisfy various quantum steering inequalities \cite{wiseprl}, of the form 
\begin{equation} \label{ineq}
\sum_j \langle a_j B_j\rangle_{\{ \rho^B_\lambda);p(\lambda\}} \leq 0 ,
\end{equation} 
where the $a_j$ denote classical random variables generated by Alice, and the $B_j$ denote quantum observables on Bob's system. To construct a quantum-refereed steering game (QRS game) from any such steering inequality, we adapt  a method recently used by Branciard {\it et al.} for constructing  games for verifying entanglement {\it per se} \cite{branc}. 

In particular, for a given steering inequality (\ref{ineq}), we define a corresponding QRS game $G$ (see Fig.~1c) in which on each run the referee, Charlie, sends Alice a classical label $j$ and Bob a state $\omega^C_k$ defined on a Hilbert space $H^C$ isomorphic to some subspace of $H^B$. These states must be such that the equivalent states $\omega^B_k$ on $H^B$ form a linear basis for the observables $B_j$, i.e., $B_j=\sum_k g_{jk}\omega^B_k$ for some set of coefficients $g_{jk}$.  Alice and Bob are not allowed to communicate during the game, but can have a prearranged strategy and perform arbitrary local operations.  Alice returns a value $a=a_j$, and Bob returns a value $b=0$ or $1$ corresponding to some POVM ${\cal B}\equiv\{ {\cal B}_0, {\cal B}_1\}$ on $H_B\otimes H_C$.  The corresponding payoff function is defined by
$P_G := \sum_{j,k} g_{jk} \langle a b\rangle_{j,k}$, 
where $\langle\cdot\rangle_{j,k}$ denotes the average over runs with a given $j$ and $k$. Alice and Bob win the game if $P_G>0$.   The QRS game in  the main text \blk is equivalent to taking $j=0,1,2,3$, $k\equiv (j,s)$, $a_j=\pm1$ for $j=1,2,3$, $a_0=-r/\sqrt{3}$,  $\omega^C_{k}=(1+s\sigma^C_j)/2$, and $g_{jk}=s$ ($=1$) for $j\neq 0$ ($j=0$).  The factor of 2  in the payoff function Eq.~(\ref{payoff}) for this game is chosen to make  $P(r)$ equal to the lefthand side of \blk the steering inequality $\sum_{j=1}^3 \langle a_j \sigma_j\rangle -r\sqrt{3}<0$  \cite{wisepra}. This steering inequality can be violated for Werner states only if $W>r/\sqrt{3}$ \cite{wisepra}, and hence this condition is also necessary for Alice and Bob to be able to win the QRS game in the main text. For perfect state generation by the referee, i.e., $r=1$ (see below), this reduces to $W>1/\sqrt{3}$ \blk

We now show Alice and Bob can win game $G$ only if  Alice and Bob share a state that is steerable by Alice.  Restricting Alice and Bob to no communication during the game prevents them from  generating a steerable state from a nonsteerable one \cite{cav}, and hence we must show that if they share any nonsteerable state on any Hilbert space $H_A\otimes H_B$ then $P_G\leq 0$.  Now,  for such a state   there is some LHS model $\{ \rho^B_\lambda;p(\lambda)\}$ (see above), and thus
\begin{eqnarray*}
P_G &=& \sum_{j,k}g_{jk}\langle  a b\rangle_{j,k} = \sum_{j,k,\lambda}g_{jk} p(\lambda)\, \langle a_j\rangle_\lambda\, {\rm Tr}_{BC}[(\rho^B_\lambda\otimes \omega^C_k) {\cal B}_1]\\
&=& \sum_{j,k,\lambda}g_{jk} N q(\lambda)\langle a_j\rangle_\lambda\, {\rm Tr}_C[\tau^C_\lambda \omega^C_k)]
=N \langle a_j B^C_j\rangle_{\{ \tau^C_\lambda;q(\lambda)\}} ,
\end{eqnarray*}
where the normalisation factor $N$, probability distribution $q(\lambda)$, and density operator $\tau^C_\lambda$ are implicitly defined via $Nq(\lambda)\tau^C_\lambda ={\rm Tr}_B[(\rho^B_\lambda\otimes \mathbbm{1}^C) {\cal B}_1]$; $B^C_j:=\sum_{k}g_{jk}\omega^C_k$ on $H_C$ is isomorphic to $B_j$ on $H_B$, and the average is with respect to the  LHS model $\{ \tau^C(\lambda);q(\lambda)\}$.  Noting the average corresponds to the left hand side of steering inequality (\ref{ineq}) for this LHS model, one has $P_G\leq 0$ as required.  Conversely, analogously to the entanglement verification games of Branciard {\it et al.}  \cite{branc},  it may be shown that Alice and Bob can in principle win the game if they share a state that violates steering inequality (\ref{ineq}), where Bob measures the projection ${\cal B}_1$ onto an appropriate Bell state on $H_B\otimes H_C$ [see, e.g., Eq.~(\ref{pwr})].

In practice, the referee cannot ensure perfect generation of the states $\omega^C_k$.  However, by performing tomography on these states, the referee can adjust the coefficients $g_{jk}$ appropriately, to take this into account.  We describe one method of doing so below, for the experiment carried out in this paper, which can be easily generalised to other QRS games. We observe that it does not matter if the generated states are acted on nontrivially by some completely positive channel, $\phi$, before reaching Bob, as this is equivalent to simply replacing Bob's measurement ${\cal B}$  on $H_B\otimes H_C$  by $(I_B\otimes\phi^*)({\cal B})$, where $\phi^*$ denotes the dual channel  and $I_B$ is the identity map on $H_B$.

In particular,  for the QRS game corresponding to Eq.~(\ref{payoff}), suppose that the referee actually generates the states  $\tilde\omega^C_{k}=\frac{1}{2}(1+{\bf n}^{(j,s)})\cdot\sigma^C$.  
The payoff function (\ref{payoff}) then evaluates to $P(r) = N\sum_\lambda q(\lambda) {\rm Tr}[\tau^C_\lambda T_\lambda(r)]$ for a shared  nonsteerable state, with $N$, $q(\lambda)$ and $\tau^C_\lambda$ defined as above and 
\begin{eqnarray*}
 T_\lambda(r) &:=& 
2\sum_j \left[\langle a_j\rangle_\lambda\left(\tilde\omega^C_{j,+}-\tilde\omega^C_{j,-}\right) -\frac{r}{\sqrt{3}} \left(\tilde\omega^C_{j,+}+\tilde\omega^C_{j,-}\right)\right]\\
&=&\left\langle \sum_j \left[ a_j\left( {\bf n}^{(j,+)} - {\bf n}^{(j,-)}\right)\right.\right. \\
&& \left.\left. ~~~~~~~~~-\frac{r}{\sqrt{3}}\left({\bf n}^{(j,+)} + {\bf n}^{(j,-)}\right)\right]\cdot\sigma^C\right\rangle_\lambda-2r\sqrt{3}\\
&\leq&  \max_{\{a_j=\pm1\}} \left| \sum_j \left[ a_j\left( {\bf n}^{(j,+)} - {\bf n}^{(j,-)}\right)\right.\right. \\
&& \left.\left. ~~~~~~~~~-\frac{r}{\sqrt{3}}\left({\bf n}^{(j,+)} + {\bf n}^{(j,-)}\right)\right]\right|-2r\sqrt{3}\\
&=& \max_{\{a_j=\pm1\}} \left| {\bf A}({\bf a})-r{\bf B}\right| - 2r\sqrt{3},
\end{eqnarray*} 
where the inequality follows using $a_j=\pm1$ and ${\bf v}\cdot\sigma\leq |{\bf v}|$, and we define ${\bf a}:=(a_1,a_2,a_3)$, ${\bf A}({\bf a}):= \sum_j  a_j\left( {\bf n}^{(j,+)} - {\bf n}^{(j,-)}\right)$, and ${\bf B}:=\sum_j\left({\bf n}^{(j,+)} + {\bf n}^{(j,-)}\right)/\sqrt{3}$. It is straightforward to show that the right hand side of the inequality is no more than zero for $r\geq r*$, with
\begin{equation}
r^*:=\max_{\{a_j=\pm1\}} \frac{\left[({\bf A}({\bf a}).{\bf B})^2 +{\bf A}({\bf a}).{\bf A}({\bf a})(3-{\bf B}.{\bf B})\right]^{1/2}-{\bf A}({\bf a}).{\bf B}}{3-{\bf B}.{\bf B}} .
\end{equation} 
Hence, for $r\geq r^*$, the operator $T_\lambda(r)$ is nonpositive, and hence $P(r)\leq0$ for any nonsteerable state.
It is straightforward to check that $r^*= 1$ for perfect state generation, $\tilde\omega^C_{k}=\omega^C_{k}=\frac{1}{2} ( \mathbbm{1}  +s\sigma_j^C)$. For the imperfect referee states generated in the experiment of this paper  we found \blk $r^*=1.081 \pm 0.009$.

\subsection*{Experimental apparatus}

The { individual} SPDC sources used in our demonstration consisted of a pair of sandwiched { Bismuth Borate (BiBO)} crystals, { each 0.5mm in length} and cut for type-I degenerate 
{ down-conversion} from 410nm (pump) to 820nm (signal/idler), with their optic axes perpendicularly oriented. Charlie's source was pumped with 200mW of horizontally polarised light to generate polarisation-unentangled photon pairs. One of Charlie's photons (signal) was sent to a 
{ single-photon counting module (Perkin-Elmer SPCM-AQR-14-FC)}, 
{ to herald the} arrival of a degenerate idler counterpart at the BSM device.  The second SPDC source was 
{ pumped} with 200mW of diagonally polarised light, 
{ generating the polarisation-entangled} state  $\rho^{AB}\neq\rho^{A}\otimes\rho^{B}$ \blk shared between Alice and Bob. The state from the SPDC source could be transformed into any of the four Bell states by implementing a local unitary with a fibre polarisation controller (to generate anti/correlated statistics)
{ combined with} a half-wave plate tilted in the $xy$ plane with its optic axis in the horizontal plane
{ (to set the} phase $\phi$ of the entangled Bell state). Alice's photon { (consisting of one-half of the entangled state)} was sent to her { single-qubit} measurement station, whereas Bob's photon { (consisting of the remaining half of the entangled state)} was coupled into single-mode fibre and sent to Bob's BSM device. Bob's BSM device consisted of a central 50:50 beam splitter and polarisation analysis at the output ports. The device combined Bob's half of the entangled state  $\rho^{AB}$, \blk and the state $\omega^C_{k}$ that Charlie sent  to \blk him.  
Bob's partial BSM device resolved the $|\Psi^{+}\rangle$ and $|\Psi^{-}\rangle$ Bell states through discrimination of orthogonally polarised photon pairs (the case of $|\Psi^{+}\rangle$) or through anti--bunching behaviour (the case of $|\Psi^{-}\rangle$). On the other hand, the $|\Phi^{\pm}\rangle$ states required number resolving detection (since these states saw pairs of photons degenerate in polarisation bunched at the point of detection). Because 
{ our single photon counting modules were not number resolving,} we instead opted for pseudo-number resolution by replacing the 
{ single-mode fibres at Bob's BSM output} with { single-mode} 50:50 fibre beam splitters. The { initially bunched} pairs of photons travelling down these { fibre beam} splitters were separated and { number-resolved} 50$\%$ of the time, a feature accounted for in the analysis of the payoff function.

The Bell state analysis
{ featured} non--classical HOM interference between the  $\rho^{{B}}$ and $\omega^C_k$ \blk photons at the central 50:50 beam splitter. 
{ A HOM interference visibility of $89\%$ was calculated, where a} high interference visibility 
{ corresponded to effective resolution of} the singlet state $|\Psi^{-}\rangle$ 
{ and the other three triplet Bell states (for some local unitary)}. Bob performed a joint measurement on $\rho^{{B}}\otimes \omega^C_k$, where the fibre input coupler for the $\omega^C_k$ photon was kept on a linear $z-$translation stage to match temporal modes between the $\rho^{{B}}$ and $\omega^C_k$ photons.
A photon detection at Alice's detector heralded the presence of the $\rho^B$ photon at the 50:50 beam splitter, and a photon detection in Charlie's  
{ heralding detector signified} the presence of the $\omega^C_k$ photon. 
Our method to calculate the payoff function $\mathrm{P}(r)$ for an experimental Werner state $\rho^{AB}$ was relatively straight--forward, and used the fact that a Werner state can be expressed as a statistical mixture of all four Bell states. Data was taken with $\rho^{AB}$ consecutively prepared in the four Bell states, and the data sets were aggregated to produce a value of the payoff function for the effective state $\rho^{AB}$. The Werner parameter was tuned by weighting the data collection time for the singlet state 
{ relative to the data} collection time for the three triplet states (where the data collection interval for the three triplet states was identical). For example, to test the payoff function using a completely mixed state ($W = 0$), data could be taken for an equal time with all four Bell states. 

Charlie's ability to send the correct state $\omega^C_k$ to Bob was also experimentally characterised. An average fidelity of $\mathcal{F}_{\mathrm{av}} = 98.7 \pm 0.6\%$ was measured in the Bell state analysis setup for the six Pauli operator eigenstates prepared by Charlie's source.

\subsection*{ Experimental error analysis}

Experimental uncertainties were derived from  Poissionian  counting statistics and standard error propagation techniques. Error bars quoted represent $\pm$ 1 standard deviation. Where uncertainties are required in quantities derived from tomographic state reconstructions \cite{white07}, the process was as follows. A large number of tomographic reconstructions  on \blk the state were performed, with each trial drawing from a  Poissonian  distribution of statistics for each measurement outcome. Each of the reconstructed density matrices were used to calculate the parameter of interest (e.g. $W$), and the mean and standard deviation of the distribution in that parameter produced the value and its uncertainty.  

\blk

}


\end{document}